\documentclass[twocolumn]{article}

\usepackage[left=1in,right=1in,top=1in,bottom=1in]{geometry}
\usepackage{tikz}
\usetikzlibrary{calc,patterns,angles,quotes}
\usetikzlibrary{arrows}
\newcommand{\trajectory}[1]{0.005*#1^2} 
\usepackage{algorithm}
\usepackage{algpseudocode}
\usepackage{amsmath}
\usepackage{authblk}
\usepackage{url}

\usepackage[explicit]{titlesec}
\renewcommand{\thesection}{\Roman{section}}
\renewcommand{\thesubsection}{\thesection.\Alph{subsection}}
\renewcommand{\thesubsubsection}{\thesubsection.\arabic{subsubsection}}

\titleformat{\section}{\large\scshape\centering}{\thesection.\space}{0pt}{#1}[]
\titlespacing*{\section}{0pt}{0.5\baselineskip}{0pt}
\titleformat{\subsection}{\normalsize\itshape}{\Alph{subsection}.\space}{0pt}{#1}[]
\titlespacing*{\subsection}{0pt}{0.5\baselineskip}{0pt}
\titleformat{\subsubsection}{\normalsize\itshape}{\arabic{subsubsection}.\space}{0pt}{#1}[]
\titlespacing*{\subsubsection}{0pt}{0.5\baselineskip}{0pt}
\usepackage{geometry}
\newgeometry{left=2cm, right=2cm, top=2.5cm,bottom=3.5cm}
\makeatletter
\renewcommand{\fnum@figure}{Fig. \thefigure}
\renewcommand{\fnum@table}{Tab. \thetable}
\makeatother

\usepackage{nopageno}
\renewcommand{\abstractname}{}

\title{\vspace{-1.0cm}\textbf{\normalsize Breaking traditions: introducing a surrogate Primer Vector in non Keplerian dynamics.}}

\author[1]{Laurent Beauregard}
\author[2]{Dario Izzo}
\author[2, 3]{Giacomo Acciarini}
\affil[1]{\emph{Telespazio GmbH, Europapl. 5, 64293 Darmstadt, servicing ESOC}}
\affil[2]{\emph{Advanced Concepts Team, European Space Research and Technology Centre (ESTEC), Noordwijk, The Netherlands.}}
\affil[3]{\emph{Surrey Space Centre, University of Surrey, Guildford, United Kingdom}}

\date{}  
\begin{document}

\maketitle

\begin{abstract}
\vspace{-1.15\baselineskip}
\textbf{\emph{\quad Abstract} - In this study, we investigate trajectories involving multiple impulses within the framework of a generic spacecraft dynamics. 
Revisiting the age-old query of \lq\lq How many impulses?\rq\rq, we present novel manipulations heavily leveraging on the properties of the state transition matrix.
Surprisingly, we are able to rediscover classical results leading to the introduction of a primer vector, albeit not making use of Pontryagin Maximum Principle as in the original developments by Lawden. 
Furthermore, our mathematical framework exhibits great flexibility and enables the introduction of what we term a \lq\lq surrogate primer vector\rq\rq\ extending a well known concept widely used in mission design. This enhancement allows to derive new simple optimality conditions that provide insights into the possibility to add and/or move multiple impulsive manoeuvres and improve the overall mass budget. This proves especially valuable in scenarios where a baseline trajectory arc is, for example, limited to a single impulse—an instance where traditional primer vector developments become singular and hinder conclusive outcomes.
In demonstrating the practical application of the surrogate primer vector, we examine a specific case involving the four-body dynamics of a spacecraft within an Earth-Moon-Sun system. The system is characterized by the high-precision and differentiable VSOP2013 and ELP2000 ephemerides models. The focal point of our investigation is a reference trajectory representing a return from Mars, utilizing the weak stability boundary (WSB) of the Sun-Earth-Moon system. The trajectory incorporates two consecutive lunar flybys to insert the spacecraft into a lunar distant retrograde orbit (DRO). Conventionally, this trajectory necessitates a single maneuver at the DRO injection point.
Prior to implementing the surrogate primer vector, a local optimization of the trajectory is performed. Upon application of the surrogate primer vector, we successfully identify potential maneuver injection points, strategically reducing the overall mission cost. The introduction of these additional maneuvers, followed by local optimization, validates that the revised trajectory indeed incurs a lower cost compared to the original configuration.}
\end{abstract}

\section{Introduction}
In this work, we will revisit the concept of the primer vector. The concept, originally developed back in the 60s by D. Lawden in a seminal work \cite{lawden1963optimal} has been more recently described in details by J. Prussing \cite{prussing2010primer} who provides a more accessible work. The original derivation uses Pontryagin optimal control theory \cite{pontryagin2018mathematical} to define an expression for the primer vector and derive necessary conditions for optimality in the context of multiple impulsive trajectories. 
In this work we propose an alternative elementary analysis of the optimality of multiple-impulse trajectory, showing how the concept of a primer vector emerges naturally without resorting to the introduction of co-states or any other concept from variational calculus (e.g. Pontryagin's theory). 

Our derivation is based on the direct analysis of the sensitivities of a multiple impulsive trajectory and provides a novel view on the fundamental concept introduced by Lawden. Using this analytical tool we are able to derive classical expressions for the primer vector for fixed time impulsive arcs \cite{lion1968primer}, as well as to extend the analysis to arcs with a single impulse, deriving new expressions for the primer vector.

\section{Context}
We consider the problem of designing optimal multi-impulsive trajectories under a generic dynamics \cite{ocampo2010variational}. While there can be multiple nonlinear constraints imposed on the spacecraft, a general statement of trajectory optimization with impulsive maneuvers can be described as follows:\\
Given a dynamical system $\bf{f}(t,x)$ and a spacecraft trajectory $\bf{x}(t)$ that obeys this dynamics
\begin{align}
\dot{\bf{x}} = \bf{f}(t,\bf{x})
\end{align}
We want to find a set of $N$ impulsive maneuvers $\Delta \bf{V}_i$ at epochs $\tau_i$ such that boundary conditions are met
\begin{align}
\bf{x}(t_s) &= \bf{x}_s\\
\bf{x}(t_f) &= \bf{x}_f
\end{align}
that also minimizes the total $\Delta v$
\begin{align}
\Delta v = \sum_i |\Delta \bf{V}_i|
\end{align}
This can be formulated in a standard nonlinear programming (NLP) problem and optimized with various algorithms such as SNOPT, WORHP, SLSQP etc. However what this formulation fails to capture is the possibility of adding new maneuvers. While it is possible to add zero velocity maneuvers a posteriori and reoptimize with these new maneuvers, the optimization result will heavily depend on the location of these new maneuvers, choosing the location poorly can results in sub-optimal configurations. Exhaustively running a search for the location of these new maneuvers can prove to be prohibitively expensive. These are the questions the primer aims to answer: Can adding a new maneuver reduce the total $\Delta v$? And if so: where should these new maneuvers be located?

\begin{figure*}[t]
\centering
\begin{tikzpicture}

\draw[blue, thin, smooth] plot[domain=0:15] ({\x},{\trajectory{\x}});


\fill[black] (0, {\trajectory{0}}) circle (1.5pt) node[below, font=\small] {$\mathbf x_s$};

\fill[black] (3.4, {\trajectory{3.4}}) circle (1.5pt) node[below left, font=\small] {$\mathbf x_1^-$} node[below right, font=\small] {$\mathbf x_1^+$} node[above left, font=\small]{$\tau_1$};
\draw[->, red] (3.4, {\trajectory{3.4}}) -- ++(95:0.9) node[above right, font=\small] {$\Delta \mathbf V_1$};
\draw [dashed, ->] (0, {\trajectory{0}}) to[out=25,in=165] node[midway, above, font=\small] {$\tilde M_s$} (3.4, {\trajectory{3.4}});

\fill[black] (6, {\trajectory{6}}) circle (1.5pt) node[below left, font=\small] {$\mathbf x_2^-$} node[below right, font=\small] {$\mathbf x_2^+$}  node[above left, font=\small]{$\tau_2$};
\draw[->, red] (6, {\trajectory{6}}) -- ++(84:0.5) node[above right, font=\small] {$\Delta \mathbf V_2$};
\draw [dashed, ->] (3.4, {\trajectory{3.4}}) to[out=25,in=165] node[midway, above, font=\small] {$\tilde M_1$} (6, {\trajectory{6}});

\fill[black] (10, {\trajectory{10}}) circle (1.5pt) node[below left, font=\small] {$\mathbf x_3^-$} node[below right, font=\small] {$\mathbf x_3^+$}  node[above left, font=\small]{$\tau_3$};
\draw[->, red] (10, {\trajectory{10}}) -- ++(60:0.6) node[above right, font=\small] {$\Delta \mathbf V_3$};
\draw [dashed, ->] (6, {\trajectory{6}}) to[out=25,in=165] node[midway, above, font=\small] {$\tilde M_2$} (10, {\trajectory{10}});

\fill[black] (15, {\trajectory{15}}) circle (1.5pt) node[below, font=\small] {$\mathbf x_f$}  node[above left, font=\small]{$\tau_f$};
\draw [dashed, ->] (10, {\trajectory{10}}) to[out=25,in=165] node[midway, above, font=\small] {$\tilde M_3$} (15, {\trajectory{15}});

\end{tikzpicture}
    \caption{A multiple impulsive trajectory with $N=3$ impulses. State $\mathbf x_i$ is considered just before the impulse is applied.}
    \label{fig:mit}
\end{figure*}
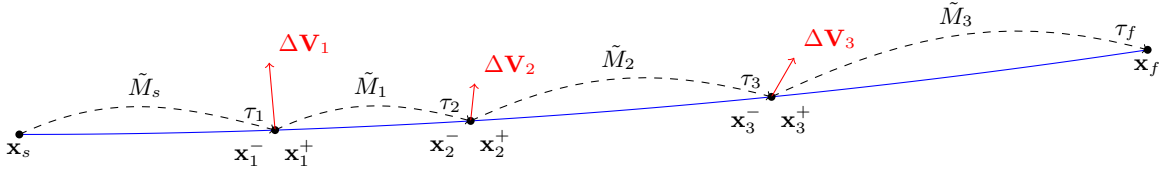

\subsection{Notation}
Before going further, we need to establish notations and conventions we are going to use throughout the paper. With reference to Fig. \ref{fig:mit} we consider a generic reference multiple-impulse trajectory we indicate with $\mathbf x_\text{ref}(t), t \in [\tau_s, \tau_f]$. The boundary states are indicated with the subscript $s$ for start and $f$ for final. The $N$ impulses are given at intermediate points at epochs $\tau_i$. Quantities immediately before the impulse are indicated with a minus (i.e. $\mathbf x^-$) and those immediately after with a $+$ (i.e. $\mathbf x^+$). The $6\times 6$ time dependent matrix which maps the perturbation $\delta \bf x_1$ of the state applied at some epoch $t_1$ to the effect $\delta \bf x_2$ on the state at a later time $t_2$ is the State Transition Matrix (STM) denoted with $M(t_1,t_2)$ or simply $M$ when possible:
\begin{equation*}
\delta \mathbf{x}_2 = M(t_1,t_2) \delta \mathbf x_1
\end{equation*}
The STM satisfies the following equations of motion:
\begin{equation*}
\dot{M} = \left. \frac{\partial \bf f}{\partial \bf x}\right|_{x_\text{ref}} M
\end{equation*}
In the case of our generic multiple impulse reference trajectory (Fig. \ref{fig:mit}), we will denote with $M_i$ the STM from $\mathbf x_j^+$ to $\mathbf x_f$. This can also be computed as the product of all intermediate STMS as follows:
$$
M_{i} = \prod_{j=i}^N {\tilde M}_j
$$
where $ {\tilde M}_j$ is the state transition matrix from  $\mathbf x_j^+$ to $\mathbf x_{j+1}^-$.

We also make heavy use of the following notation in manipulations involving STMs:
\begin{equation*}
M = \left[ 
\begin{array}{c|c} 
  M^{rr} & M^{rv} \\ 
  \hline 
  M^{vr} & M^{vv} 
\end{array} 
\right] 
= \left[ 
\begin{array}{c|c} 
  M^{xr} & M^{xv} 
\end{array} 
\right] 
= \left[ 
\begin{array}{c} 
  M^{rx} \\
  \hline 
  M^{vx} 
\end{array} 
\right] 
\end{equation*}
so that we may write, for example:
\begin{subequations}
\begin{align}
\delta \mathbf r_f = M^{rr}\delta \mathbf r_s + M^{rv}\delta \mathbf v_s\\
\delta \mathbf v_f = M^{vr}\delta \mathbf r_s + M^{vv}\delta \mathbf v_s
\end{align}
\end{subequations}
as well as the more compact form:
$$
\delta  \mathbf x_f = M(\tau_s, \tau_f)\delta  \mathbf x_s
$$

\subsection{Sensitivities in a multi-impulsive trajectory}
A generic multiple-impulse trajectory is fully defined by the initial state, the dynamics and the various $\Delta \bf{V}_i$ applied at $\tau_i$. The differential of the final state with respect to variations of the free parameters is:
\begin{align}
\label{eq:state_variation}
\delta \mathbf x_f &= \sum_{i=1}^N M_{i}^{xv}\delta \Delta \mathbf V_i + \sum_{i=1}^N M_{i}\left( \mathbf f_i^- -\mathbf f_{i}^+\right)\delta \tau_i
\end{align}
with
\begin{align}
M_i = M(\tau_i,\tau_f)
\end{align}
as it can be readily seen applying the definition of the STM and singling out the various contributions of the resulting state changes.
Whenever the dynamics has the form $\bf{f} = [v,g(t,r)]$, then the previous formula simplifies to:
\begin{align}
\delta \mathbf x_f &= \sum_{i=1}^N M_{i}^{xv}\delta \Delta \mathbf V_i - \sum_{i=1}^N M_{i}^{xr} \Delta \mathbf V_i\delta \tau_i
\end{align}
This form is essentially widely applicable unless there is a force that depends on velocity (e.g. a drag term, a rotating frame, etc.). 
When designing a multiple impulsive trajectory, it is often the case that the sum of all velocity increment magnitudes $\Delta v = \sum_i |\Delta \mathbf V_i|$ is considered as an objective function to minimize. 
Let us now assume that we are aiming to lower the total $\Delta v$ by adding a number of infinitesimally small velocity increments to the already existing impulses, as well as some additional ones at some other epochs (i.e. in correspondence to zero norm impulses). 

The total variation of the objective function can then be expressed in the following general form:
\begin{align}
\label{eq:objective}
\delta \Delta v = \sum_{|\Delta \mathbf V_i|\neq 0} \frac{\Delta \mathbf{V}_i}{|\Delta \mathbf{V}_i|}\cdot \delta \Delta \mathbf{V}_i+\sum_{|\Delta \mathbf V_i| = 0} |\delta \Delta \mathbf{V}_i|
\end{align}
Note how the impulses added in correspondence of the finite $\Delta \mathbf V_i$ have a much different effect on the total $\Delta v$ variation. This, as we shall see, has profound consequences in the optimality conditions we will derive.

\section{Adding/changing impulses}
To simplify the initial analysis we will consider the possibility to  add 3 correction maneuvers $\delta \Delta \mathbf V_i$ at the epochs $\tau_i$ to an already existing multiple impulsive arc. Later we will see that this, in fact, cover all possible cases. The three correction maneuvers (velocity increments) are applied at epochs where the impulses $\Delta \mathbf V_1$, $\Delta \mathbf V_2$, $\Delta \mathbf V_3$ are present. We allow the possibility that some of those $\Delta \mathbf V_i$ are actually zero, in which case the correction manoeuvre corresponds to the addition of one impulse. Without loss of generality, we impose the order that $\Delta \mathbf V_1$ occurs at the \textbf{beginning}  of the trajectory, $\Delta \mathbf V_3$ occurs at the \textbf{end} of the segment and that $\Delta \mathbf V_2$ lies in between. The total $\Delta v$, before applying the correction manoeuvres, is given by:
\begin{align}
    \Delta v = |\Delta \mathbf V_1|+|\Delta \mathbf V_2|+|\Delta \mathbf V_3|
\end{align}
after:
\begin{align}
    \Delta v = |\Delta \mathbf V_1+\delta \Delta \mathbf V_1|+|\Delta \mathbf V_2+\delta \Delta \mathbf V_2|+|\Delta \mathbf V_3+\delta \Delta \mathbf V_3|
\end{align}
Is it possible to find the $\delta \Delta \mathbf V$ as to reduce the total $\Delta v$ while still reaching the very same final conditions? This is the fundamental question we will be answering definitively in the rest of our developments, and that ties closely to Edelbaum's question\lq\lq How many impulses?\rq\rq\cite{edelbaum1967many}.
Since we need to reach the same end state, applying Eq.(\ref{eq:state_variation}), the boundary conditions differentials can be written as:
\begin{subequations}
\begin{align}
\delta \mathbf r_f &= M_1^{rv}\delta \Delta \mathbf V_1 + M_2^{rv}\delta \Delta \mathbf V_2 = 0\\
\delta \mathbf v_f &= M_1^{vv}\delta \Delta \mathbf V_1 + M_2^{vv}\delta \Delta \mathbf V_2 +\delta \Delta \mathbf V_3 = 0
\end{align}
\end{subequations}
Where
\begin{align*}
M_1 &= M(\tau_1,\tau_3)\\
M_2 &= M(\tau_2,\tau_3)
\end{align*}
Solving for $\delta \Delta \mathbf V_1$ in terms of $\delta\Delta \mathbf V_2$, we obtain
\begin{align}
    \delta\Delta \mathbf V_1 = \left[\left(M_1^{rv}\right)^{-1}M_2^{rv}\right]\delta \Delta \mathbf V_2 = A_{12}\delta \Delta \mathbf V_2
\end{align}
According to Prussing \cite{prussing2010primer} the set of times for which $M_1^{rv}$ is not invertible is discrete and so we can neglect these cases. Now $\delta \Delta \mathbf V_3$ can be solved in terms of $\delta \Delta \mathbf V_2$
\begin{align}
    \delta \Delta \mathbf V_3 = \left[-M_1^{vv}A_{12}-M_2^{vv}\right]\delta \Delta \mathbf V_2
\end{align}
For ease of notation, we will rewrite those equations in the following way:
\begin{subequations}
\label{dv_form}
\begin{align}
    \delta\Delta \mathbf V_1 &= A_{12}\delta \Delta \mathbf V_2\\
        \delta \Delta \mathbf V_3 &= A_{32}\delta \Delta \mathbf V_2
\end{align}
\end{subequations}
with
\begin{subequations}
\label{submatrix}
\begin{align}
    A_{12} &= \left(M_1^{rv}\right)^{-1}M_2^{rv}\\
    A_{32} &= -M_1^{vv}A_{12}-M_2^{vv}
\end{align}
\end{subequations}
These linear relations, efficiently computed from the state transition matrix (available via the variational equations or even analytically for simpler dynamics), introduce a relation between the three correction manoeuvres, so that we are only free to decide the middle one (or any other) as the other two will follow necessarily as to satisfy the requested boundary conditions.
In order to search for the existence of a $\delta \Delta \mathbf V_2$ that lowers the total objective, and with reference to Eq.(\ref{eq:objective}), there are 4 qualitatively different cases which is interesting to analyze here (there are more in the most general case, but we here restrict ourselves to these), as seen in table \ref{Cases}:

\begin{table}[]
\centering
\caption{Case taxonomy \label{Cases}}
\begin{tabular}{|c|c|c|l|}
\hline
& \multicolumn{2}{c|}{\#$\Delta \mathbf V$} & \\

Case & $=0$ & $\ne 0 $ & Comment                 \\ \hline
1           & 3                  & 0                    & Trivial case            \\ \hline
2           & 0                  & 3                    & Local optimality        \\ \hline
3           & 1                  & 2                    & Primer vector \\ \hline
4           & 2                  & 1                    & Surrogate primer vector       \\ \hline
\end{tabular}
\end{table}

\subsection{Case 1: 3 zero $\Delta \mathbf V$'s}
The first case of 3 zero $\Delta \mathbf V$ is trivial as adding any new $\Delta \mathbf V$ will automatically increase the total $\Delta v$
\subsection{Case 2: 3 nonzero $\Delta \mathbf V$'s}
The most straightforward case is 3 nonzero $\Delta \mathbf V$'s -- Eq.(\ref{eq:objective}) reduces to:
\begin{align}
\delta \Delta v = \frac{\Delta \mathbf V_1}{|\Delta \mathbf V_1|}\cdot\delta\Delta \mathbf V_1+\frac{\Delta \mathbf V_2}{|\Delta \mathbf V_2|}\cdot\delta\Delta \mathbf V_2+\frac{\Delta \mathbf V_3}{|\Delta \mathbf V_3|}\cdot\delta\Delta \mathbf V_3
\end{align}
This can be written in terms of $\delta \Delta \mathbf V_2$ only
\begin{align}
\delta \Delta v = \left(\frac{\Delta \mathbf V_1}{|\Delta \mathbf V_1|}\cdot A_{12}+\frac{\Delta \mathbf V_2}{|\Delta \mathbf V_2|}+\frac{\Delta \mathbf V_3}{|\Delta \mathbf V_3|}\cdot A_{32}\right)\cdot \delta \Delta \mathbf V_2
\end{align}
The total $\Delta v$ can be reduced if the vector $\mathbf b$ is nonzero, where $\mathbf b$ is
\begin{align}
\mathbf b = -\left(\frac{\Delta \mathbf V_1}{|\Delta \mathbf V_1|}\cdot A_{12}+\frac{\Delta \mathbf V_2}{|\Delta \mathbf V_2|}+\frac{\Delta \mathbf V_3}{|\Delta \mathbf V_3|}\cdot A_{32}\right)
\end{align}
In the case that $|\mathbf b|>0$ we can choose
\begin{align}
    \delta \Delta \mathbf V_2 &= |\delta \Delta \mathbf V_2| \frac{\mathbf b}{|\mathbf b|}
\end{align}
And $\delta \Delta \mathbf V_1$, $\delta \Delta \mathbf V_3$ from equations \ref{dv_form}. These directions will guarantee to locally reduce the total $\Delta v$. If the vector $\mathbf b$ is 0 then we know, at least locally, the total $\Delta v$ cannot be reduced
\subsection{Case 3: 2 nonzero $\Delta \mathbf V$'s, 1 zero $\Delta \mathbf V$}
Assuming $\delta \Delta \mathbf V_2 = \mathbf 0$, this is the case that will result in finding the classical primer vector -- Eq.(\ref{eq:objective}) reduces to:
\begin{align}
\delta \Delta v = \frac{\Delta \mathbf V_1}{|\Delta \mathbf V_1|}\cdot\delta\Delta \mathbf V_1+|\delta\Delta \mathbf V_2|+\frac{\Delta \mathbf V_3}{|\Delta \mathbf V_3|}\cdot\delta\Delta \mathbf V_3
\end{align}
Writing everything in terms of $\delta\Delta \mathbf V_2$
\begin{align}
\delta \Delta v = \left[\frac{\Delta \mathbf V_1}{|\Delta \mathbf V_1|}\cdot A_{12}+\frac{\Delta \mathbf V_3}{|\Delta \mathbf V_3|}\cdot A_{32}\right]\cdot\delta\Delta \mathbf V_2+|\delta\Delta \mathbf V_2|
\end{align}
Letting $\delta\Delta \mathbf V_2 = |\delta\Delta \mathbf V_2| \hat{\mathbf u}$ where $\hat{\mathbf u}$ is a unit vector, allows us to write
\begin{align}
\delta \Delta v = \left(\left[\frac{\Delta \mathbf V_1}{|\Delta \mathbf V_1|}\cdot A_{12}+\frac{\Delta \mathbf V_3}{|\Delta \mathbf V_3|}\cdot A_{32}\right]\cdot\hat{\mathbf u}+1\right)|\delta\Delta \mathbf V_2|
\end{align}

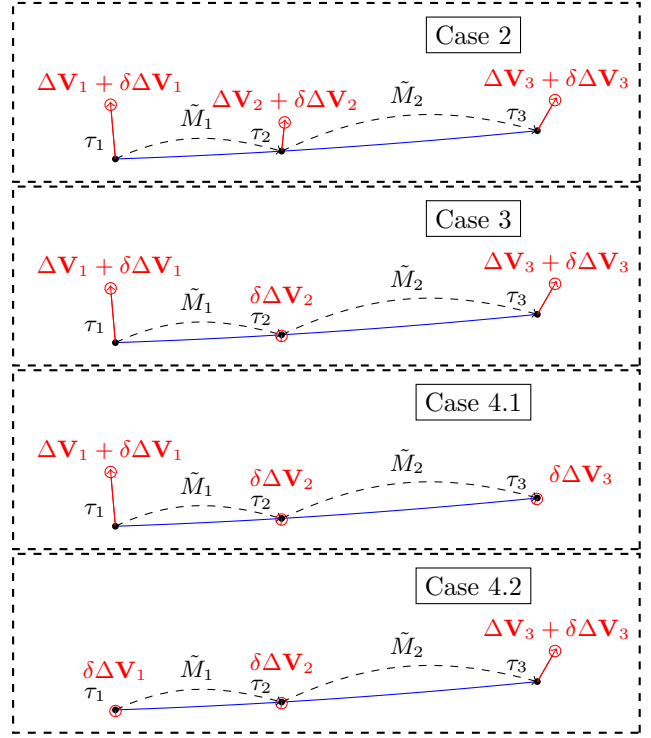
\begin{figure}[t]
\centering
\begin{tikzpicture}[scale=0.85]
\draw[dashed,thick] (11.6,-0.3) rectangle (1.8,2.5);
\node[draw] at (9,2) {Case 2};
\draw[blue, thin, smooth] plot[domain=3.4:10] ({\x},{\trajectory{\x}});
\fill[black] (3.4, {\trajectory{3.4}}) circle (1.5pt) node[above left, font=\small] {$\tau_1$};
\draw[->, red] (3.4, {\trajectory{3.4}}) -- ++(95:0.9) node[above, font=\small] {$\Delta \mathbf V_1 + \delta \Delta \mathbf V_1 $};
\draw[red, -o] (3.4, {\trajectory{3.4}}) -- ++(95:0.95);
\draw [dashed, ->] (3.4, {\trajectory{3.4}}) to[out=25,in=165] node[midway, above, font=\small] {$\tilde M_1$} (6, {\trajectory{6}});
\fill[black] (6, {\trajectory{6}}) circle (1.5pt) node[above left, font=\small] {$\tau_2$};
\draw[->, red] (6, {\trajectory{6}}) -- ++(84:0.5) node[above, font=\small] {$\Delta \mathbf V_2+ \delta \Delta \mathbf V_2 $};
\draw[red, -o] (6, {\trajectory{6}}) -- ++(84:0.55);
\fill[black] (10, {\trajectory{10}}) circle (1.5pt) node[above left, font=\small] {$\tau_3$};
\draw[->, red] (10, {\trajectory{10}}) -- ++(60:0.6) node[above, font=\small] {$\Delta \mathbf V_3+ \delta \Delta \mathbf V_3 $};
\draw[red, -o] (10, {\trajectory{10}}) -- ++(60:0.65);
\draw [dashed, ->] (6, {\trajectory{6}}) to[out=25,in=165] node[midway, above, font=\small] {$\tilde M_2$} (10, {\trajectory{10}});
\end{tikzpicture}

\begin{tikzpicture}[scale=0.85]
\draw[dashed,thick] (11.6,-0.3) rectangle (1.8,2.5);
\node[draw] at (9,2) {Case 3};
\draw[blue, thin, smooth] plot[domain=3.4:10] ({\x},{\trajectory{\x}});
\fill[black] (3.4, {\trajectory{3.4}}) circle (1.5pt) node[above left, font=\small] {$\tau_1$};
\draw[->, red] (3.4, {\trajectory{3.4}}) -- ++(95:0.9) node[above, font=\small] {$\Delta \mathbf V_1 + \delta \Delta \mathbf V_1 $};
\draw[red, -o] (3.4, {\trajectory{3.4}}) -- ++(95:0.95);
\draw [dashed, ->] (3.4, {\trajectory{3.4}}) to[out=25,in=165] node[midway, above, font=\small] {$\tilde M_1$} (6, {\trajectory{6}});
\fill[black] (6, {\trajectory{6}}) circle (1.5pt) node[above left, font=\small] {$\tau_2$};
\draw[red, -o] (6, {\trajectory{6}}) -- ++(84:0.08) node[above=.2cm, font=\small] {$\delta \Delta \mathbf V_2 $};
\fill[black] (10, {\trajectory{10}}) circle (1.5pt) node[above left, font=\small] {$\tau_3$};
\draw[->, red] (10, {\trajectory{10}}) -- ++(60:0.6) node[above, font=\small] {$\Delta \mathbf V_3+ \delta \Delta \mathbf V_3 $};
\draw[red, -o] (10, {\trajectory{10}}) -- ++(60:0.65);
\draw [dashed, ->] (6, {\trajectory{6}}) to[out=25,in=165] node[midway, above, font=\small] {$\tilde M_2$} (10, {\trajectory{10}});
\end{tikzpicture}

\begin{tikzpicture}[scale=0.85]
\draw[dashed,thick] (11.6,-0.3) rectangle (1.8,2.5);
\node[draw] at (9,2) {Case 4.1};
\draw[blue, thin, smooth] plot[domain=3.4:10] ({\x},{\trajectory{\x}});
\fill[black] (3.4, {\trajectory{3.4}}) circle (1.5pt) node[above left, font=\small] {$\tau_1$};
\draw[->, red] (3.4, {\trajectory{3.4}}) -- ++(95:0.9) node[above, font=\small] {$\Delta \mathbf V_1 + \delta \Delta \mathbf V_1 $};
\draw[red, -o] (3.4, {\trajectory{3.4}}) -- ++(95:0.95);
\draw [dashed, ->] (3.4, {\trajectory{3.4}}) to[out=25,in=165] node[midway, above, font=\small] {$\tilde M_1$} (6, {\trajectory{6}});
\fill[black] (6, {\trajectory{6}}) circle (1.5pt) node[above left, font=\small] {$\tau_2$};
\draw[red, -o] (6, {\trajectory{6}}) -- ++(84:0.08) node[above=0.2cm, font=\small] {$\delta \Delta \mathbf V_2 $};
\fill[black] (10, {\trajectory{10}}) circle (1.5pt) node[above left, font=\small] {$\tau_3$};
\draw[red, -o] (10, {\trajectory{10}}) -- ++(60:0.08) node[above right, font=\small] {$\delta \Delta \mathbf V_3 $};
\draw [dashed, ->] (6, {\trajectory{6}}) to[out=25,in=165] node[midway, above, font=\small] {$\tilde M_2$} (10, {\trajectory{10}});
\end{tikzpicture}

\begin{tikzpicture}[scale=0.85]
\draw[dashed,thick] (11.6,-0.3) rectangle (1.8,2.5);
\node[draw] at (9,2) {Case 4.2};
\draw[blue, thin, smooth] plot[domain=3.4:10] ({\x},{\trajectory{\x}});
\fill[black] (3.4, {\trajectory{3.4}}) circle (1.5pt) node[above left, font=\small] {$\tau_1$};
\draw[red, -o] (3.4, {\trajectory{3.4}}) -- ++(95:0.08) node[above=0.2cm, font=\small] {$\delta \Delta \mathbf V_1 $};
\fill[black] (6, {\trajectory{6}}) circle (1.5pt) node[above left, font=\small] {$\tau_2$};
\draw[red, -o] (6, {\trajectory{6}}) -- ++(84:0.08) node[above=0.2cm, font=\small] {$\delta \Delta \mathbf V_2 $};
\draw [dashed, ->] (3.4, {\trajectory{3.4}}) to[out=25,in=165] node[midway, above, font=\small] {$\tilde M_1$} (6, {\trajectory{6}});
\fill[black] (10, {\trajectory{10}}) circle (1.5pt) node[above left, font=\small] {$\tau_3$};
\draw[->, red] (10, {\trajectory{10}}) -- ++(60:0.6) node[above, font=\small] {$\Delta \mathbf V_3+ \delta \Delta \mathbf V_3 $};
\draw[red, -o] (10, {\trajectory{10}}) -- ++(60:0.65);
\draw [dashed, ->] (6, {\trajectory{6}}) to[out=25,in=165] node[midway, above, font=\small] {$\tilde M_2$} (10, {\trajectory{10}});
\end{tikzpicture}

\caption{Non trivial cases considered. From Case 2 we find local optimality conditions. From Case 3 we are able to derive an expression for the primer vector. From Case 4 we find a new necessary condition for optimality allowing, also for this case where a primer vector is not available, to conclude on the possibility to add impulses.}
\label{fig2:mit}
\end{figure}

Let us introduce the vector $\mathbf p$ defined as:
\begin{equation}
\label{eq:primer}
   \mathbf p = -\left(\frac{\Delta \mathbf V_1}{|\Delta \mathbf V_1|}\cdot A_{12}+\frac{\Delta \mathbf V_3}{|\Delta \mathbf V_3|}\cdot A_{32}\right)
\end{equation}
We get:
\begin{align}
\delta \Delta v = \left(-\mathbf p\cdot\hat{\mathbf u}+1\right)|\delta\Delta \mathbf V_2|
\end{align}
The best direction to lower the total $\Delta v$ is $\hat{\mathbf u} = \mathbf p/|\mathbf p|$, in which case the total reduction of $\Delta v$ is given by
\begin{align}
\delta \Delta v = \left(-|\mathbf p|+1\right)|\delta\Delta \mathbf V_2|
\end{align}
This means if $|\mathbf p|>1$ there exist a direction that can lower the total $\Delta v$. Note that Eq. (\ref{eq:primer}), accounting for Eq.(\ref{submatrix}), is the same as Eq.(2.50) in \cite{prussing2010primer}. We have thus found, in this case, an expression for the \emph{primer vector}, remarkably circumventing entirely Pontryagin's theory and without introducing co-states. 
The corresponding change in the $\Delta \mathbf V$'s are given by
\begin{subequations}
\begin{align}
    \delta\Delta \mathbf V_2 &= |\delta\Delta \mathbf V_2|\frac{\mathbf p}{|\mathbf p|}
\end{align}
\end{subequations}
And $\delta \Delta \mathbf V_1$, $\delta \Delta \mathbf V_3$ from equations \ref{dv_form}. If everywhere on the trajectory $|\mathbf p|\leq 1$ then, locally, it is not possible to add a new maneuver and reduce the total $\Delta v$.

\subsection{Case 4: 1 nonzero $\Delta \mathbf V$, 2 zero $\Delta \mathbf V$'s}
This is a novel application of the primer vector. Here we have only 1 non-zero $\Delta \mathbf V$. We can still solve for $\delta \Delta \mathbf V_1$ and $\delta \Delta \mathbf V_3$ in terms of $\delta \Delta \mathbf V_2$ with the formula \ref{dv_form} and \ref{submatrix}.
\subsubsection{$\Delta \mathbf V_1 \neq 0$, $\Delta \mathbf V_3 = 0$}
Eq.(\ref{eq:objective}) reduces to:
\begin{align}
\delta \Delta v = \frac{\Delta \mathbf V_1}{|\Delta \mathbf V_1|}\cdot\delta\Delta \mathbf V_1+|\delta \Delta \mathbf V_3|+|\delta \Delta \mathbf V_2|
\end{align}
Writing everything in terms of $\delta \Delta \mathbf V_2$
\begin{align}
\delta \Delta v = \frac{\Delta \mathbf V_1}{|\Delta \mathbf V_1|}\cdot A_{12}\cdot\delta \Delta \mathbf V_2+|A_{32}\cdot\delta \Delta \mathbf V_2|+|\delta \Delta \mathbf V_2|
\end{align}
Similarly to what done previously we have:
\begin{align}
\delta \Delta v = \left(\left[\frac{\Delta \mathbf V_1}{|\Delta \mathbf V_1|}\cdot A_{12}\right]\cdot\hat{\mathbf u}+|A_{32}\cdot\hat{\mathbf u}|+1\right)|\delta \Delta \mathbf V_2|
\end{align}
where 
\begin{align}
    \mathbf b = -\left(\frac{\Delta \mathbf V_1}{|\Delta \mathbf V_1|}\cdot A_{12}\right)
\end{align}
and $\delta \Delta \mathbf V_2 = |\delta \Delta \mathbf V_2|\hat{\mathbf u}$.

\subsubsection{Case $\Delta \mathbf V_1=0$, $\Delta \mathbf V_3 \neq 0$}
This case result in the same developments as in the previous case where the indexes are swapped $1 \longleftrightarrow 3$.\\\\
In all cases, an expression in the following form is found:
\begin{align}
\delta \Delta v = \left(-\mathbf b\cdot\hat{\mathbf u}+|B\hat{\mathbf u}|+1\right)|\delta \Delta \mathbf V_2|
\end{align}
We may then obtain a reduction in our total $\Delta v$ if and only if $\min_{\hat{\mathbf u}}\left(-\mathbf b\cdot\hat{\mathbf u}+|B\hat{\mathbf u}|+1\right) < 0$.
Trivially, if $|\mathbf b|<1$ then we can exclude such a possibility efficiently avoiding to make any computation. Besides, since:
\begin{align}
    \delta \Delta v\geq \left(-|\mathbf b| + \sigma_\text{min} + 1)\right)|\delta \Delta \mathbf V_2|
\end{align}
where $\sigma_\text{min}$ is the smallest singular value of the matrix $B$ we have that if
\begin{align}
|\mathbf b| < 1 + \sigma_\text{min}
\end{align}
we can also exclude to be able to reduce the total $\Delta V$. In other cases, we must find numerically the minimum. There are 2 good initial guesses to the minimization problem.
\begin{itemize}
    \item $\hat{\mathbf u} = \mathbf b/|\mathbf b|$
    \item $\hat{\mathbf u} = \pm \mathbf s_\text{min}$
\end{itemize}
where $\mathbf s_\text{min}$ is the singular vector that corresponds to the smallest singular value of the matrix $B$. The sign in front of $\mathbf s_\text{min}$ is chosen so that
\begin{align}
    \hat{\mathbf u}\cdot \mathbf b \geq 0
\end{align}
With these 2 initial guesses, numerical analysis indicates the global minima of the function is always obtain using a few iterations of a gradient based solver. Once this optimization is done and $\hat{\mathbf u}_\text{optimal}$ is obtained, we need to check if the \textbf{primer condition} is satisfied
\begin{align}
\mathbf b\cdot\hat{\mathbf u}_\text{optimal}-|B\hat{\mathbf u}_\text{optimal}|>1
\end{align}
In case the condition is satisfied, the directions that produces a lower total $\Delta v$ are given by
\begin{align}
    \delta \Delta \mathbf V_2 &= |\delta \Delta \mathbf V_2|\hat{\mathbf u}_\text{optimal}
\end{align}
And $\delta \Delta \mathbf V_1$, $\delta \Delta \mathbf V_3$ from equations \ref{dv_form}.
To be consistent with the classical primer case, it may be useful to define a \textbf{surrogate primer vector} $\mathbf p$ as: 
\begin{align}
\mathbf p = \Big(\mathbf b\cdot\hat{\mathbf u}_\text{optimal}-|B\hat{\mathbf u}_\text{optimal}|\Big)\hat{\mathbf u}_\text{optimal}
\end{align}
\subsection{Extra commentary}
Please note that when one of the $\Delta \mathbf V = 0$, its location is arbitrary along the trajectory, so the above computation should be done for each possible location (Or at least densely enough). For case 3), there is one zero $\Delta \mathbf V$ so the primer vector is parameterized by the location of the new maneuver along the orbit. In case 4) the novel application, there are 2 zero maneuvers. This means the analysis must be performed for all possible pairs of time $t_1<t_2$ along the trajectory.
\subsection{Independence of the primer vector}
In the above analysis, we considered the case of only 3 maneuvers. However, there are trajectories with arbitrary many maneuvers. How does the primer vector behave in such a case? Concentrating on the classical primer vector case: at any given time along the trajectory, the primer vector can be calculated by choosing any two nonzero maneuvers. In general, the primer vector will be dependent of the choice of the two maneuvers. However, we have shown that if the trajectory is locally optimal with respect to all $\Delta \mathbf V_i$ then the primer vector is independent of the choice of the two maneuvers chosen to calculate it. In fact, if the trajectory is locally optimal, the primer vector, at any time, has a simple form
\begin{align}
\mathbf p(t) = \boldsymbol\lambda_r\cdot M(t,t_f)^{rv}+
\boldsymbol\lambda_v\cdot M(t,t_f)^{vv}
\end{align}
where $\boldsymbol\lambda_r$ and $\boldsymbol\lambda_v$ are constant vectors, their values are
\begin{subequations}
 \begin{align}
\boldsymbol\lambda_v &= \frac{\Delta \mathbf V_f}{|\Delta \mathbf V_f|}\\
\boldsymbol\lambda_r &= \left(\frac{\Delta \mathbf V_s}{|\Delta \mathbf V_s|}-\boldsymbol\lambda_v\cdot M(t_s,t_f)^{vv}\right)\Big[M(t_s,t_f)^{rv}\Big]^{-1}
\end{align}
\end{subequations}
$\Delta \textbf V_s$ and $\Delta \textbf V_f$ are the first and last nonzero maneuver respectively. Interestingly, only optimality with respect to $\Delta \mathbf V_i$ and not the epoch $\tau_i$ is needed. The proof of independence, while not complicated, is a bit tedious and so is omitted.

\begin{figure}
    \centering
    \includegraphics[width=0.8\columnwidth]{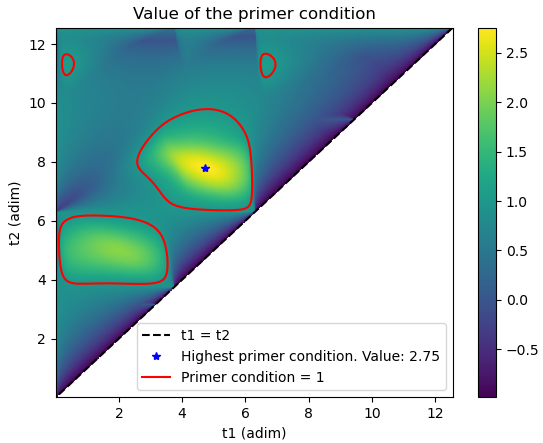}
    \caption{Surrogate primer vector map for the simple transfer}
    \label{primer_example}
\end{figure}

\section{Examples}
\subsection{A simple transfer}
\begin{figure}[tb]
    \centering
    \includegraphics[width=0.8\columnwidth]{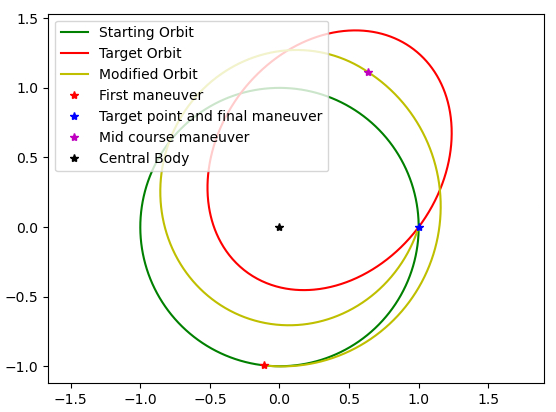}
    \caption{Example of the change in the trajectory following surrogate primer vector for the simple transfer.}
    \label{trajectory_example}
\end{figure}

In this chapter, we will cover a simple application of the novel primer condition in a Keplerian dynamics transfer. Although the case itself is trivial, it serves as a toy model that can be well understood to illustrate the concept. The system in question will be scaled so that the center body has $\mu = GM = 1$. The transfer start with the state
\begin{align}
r_0 &= [1,0,0]\\
v_0 &= [0,1,0]
\end{align}
The spacecraft is propagated for $4\pi$ units of time (and reaches the same state) and then the maneuver
\begin{align}
\Delta \mathbf V_3 &= [0.6,-0.2,0]
\end{align}
is applied. The transfer goes from a circular orbit of semi-major of 1 to an eccentric orbit of the same semi-major axis of 1. This transfer cost $\Delta v_\text{total} = \sqrt{2/5} \approx 0.632$. Applying the surrogate primer vector calculation for $0\leq t_1\leq t_2\leq 4\pi$, we obtain the graph in \ref{primer_example}.
The maximum of the value of
\begin{align}
\mathbf b\cdot\hat{\mathbf u}_\text{optimal}-|B\hat{\mathbf u}_\text{optimal}|
\end{align}
is reached at $t_1 = 4.708$ and $t_2 = 7.783$ with value $2.754$. The corresponding $\delta \Delta \mathbf V's$ are
\begin{align*}
\delta \Delta \mathbf V_1/|\delta \Delta \mathbf V_2| &= [ 0.941, 0.036 , 0 ]\\
\delta \Delta \mathbf V_2/|\delta \Delta \mathbf V_2| &= [ 0.997 , -0.078 , 0 ]\\
\delta \Delta \mathbf V_3/|\delta \Delta \mathbf V_2| &= [ -3.878 , 0.05834 ,0 ]
\end{align*}
An example of the trajectory that is obtained following the direction of the surrogate primer vector is shown in \ref{trajectory_example}. The trajectory shown has a $\Delta v_\text{total} = 0.487$

\subsection{The MSR-ERO case}
We now turn our attention to a more operational case. In the context of the MSR-ERO mission \cite{you2023mars}, one of the proposed trajectories for the Earth Return Orbiter (ERO), coming back from Mars makes use of a lunar fly to insert into the weak stability boundary \cite{garcia2007note} of the Earth-Moon-Sun system and then a second flyby of the Moon to insert in a Distant Retrograde Orbit (DRO) \cite{capdevila2018transfer} of the Earth-Moon system. This trajectory has nominally only 1 maneuver, a burn at the insertion into the DRO. The total $\Delta v$ for this trajectory is 72.89 m/s. We reproduced the trajectory under a dynamics accounting only for the Earth, the Sun and the Moon gravity using the VSOP and ELP analytical ephemerides in a Taylor integration scheme \cite{biscani2021revisiting}.

\begin{figure}[tb]
    \centering
    \includegraphics[width=0.8\columnwidth]{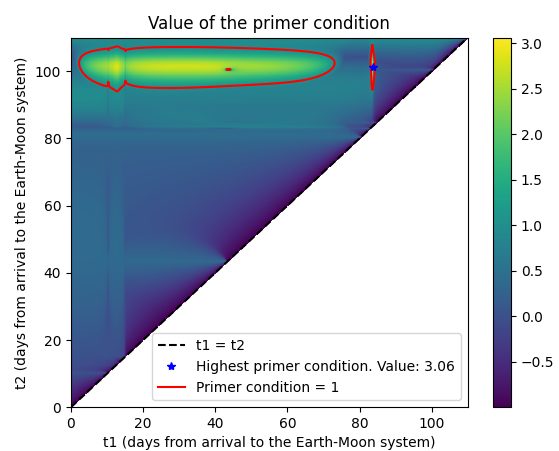}
    \caption{Surrogate primer vector map for the MSR-ERO case}
    \label{msr_ero_primer}
\end{figure}

\begin{figure}
    \centering
    \includegraphics[width=0.8\columnwidth]{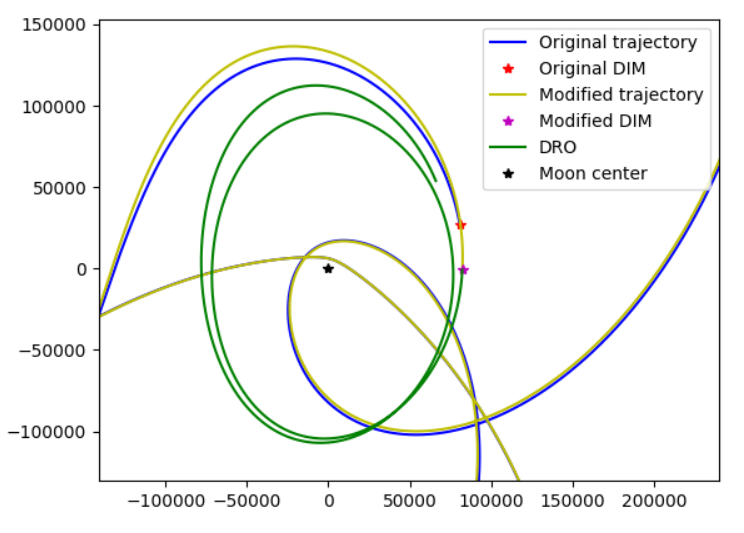}
    \caption{End part of the trajectories inserting into the DRO. The trajectories are displayed in the Earth-Moon rotating system, centered on the Moon. For the modified trajectory the other maneuvers are outside this plot.}
    \label{msr_ero_traj}
\end{figure}

The surrogate primer vector was calculated and is shown in figure \ref{msr_ero_primer}. Several locations exist that can lower the total $\Delta v$. Numerical optimization of the trajectory using these locations shows the $\Delta v$ can be reduced by $\approx $ 3 m/s. While this is a modest reduction in $\Delta v$, this case shows the possibility and general applicability of our method to generic multiple impulse trajectories. The resulting trajectories are shown in \ref{msr_ero_traj}.
[tb]

\begin{figure}[t]
    \centering
    \includegraphics[width=0.9\columnwidth]{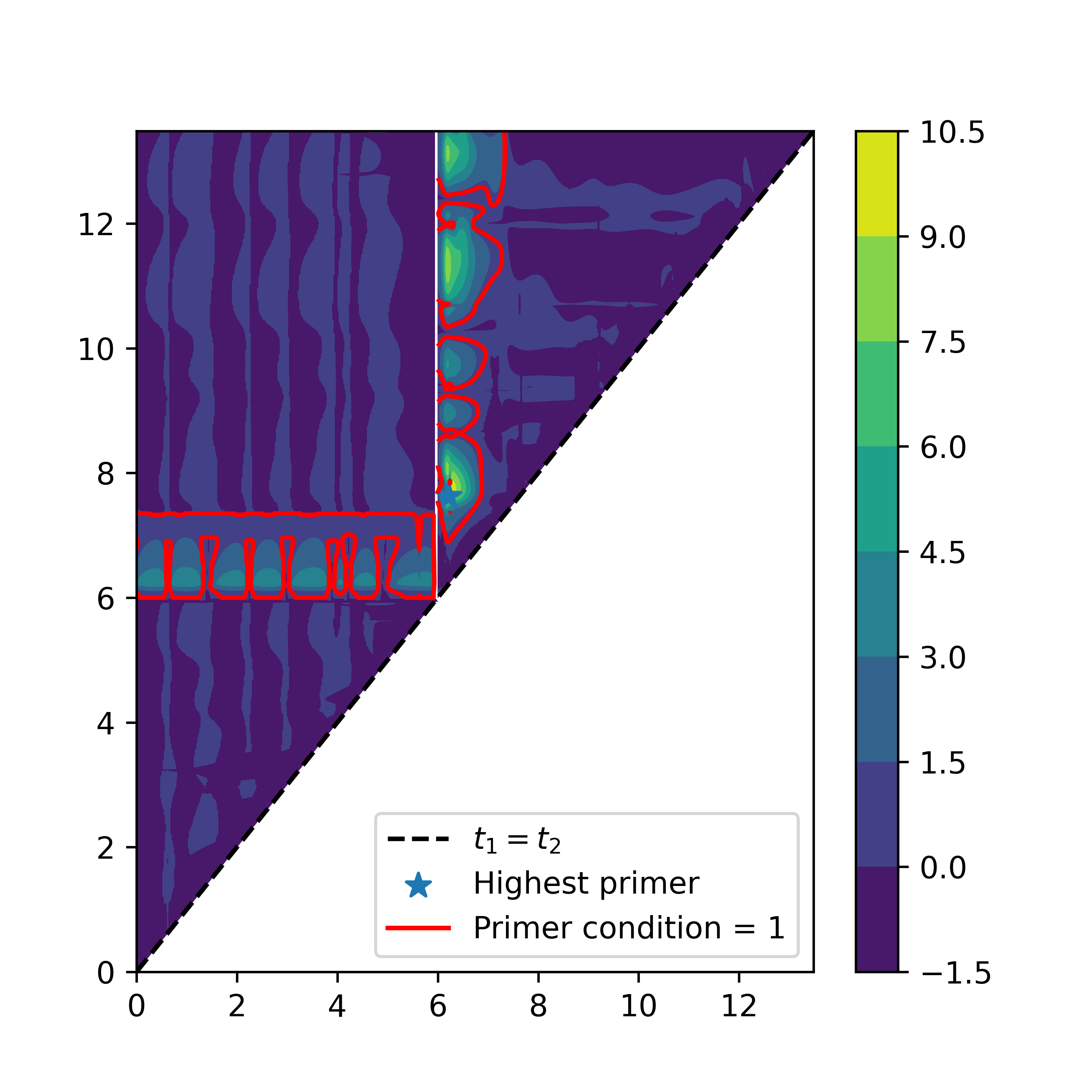}
    \caption{Surrogate primer vector map for the Halo-Vertical Lyapunov case.}
    \label{fig:heteroclinic_s}
\end{figure}

\begin{figure*}[ht]
    \centering
    \includegraphics[width=0.9\textwidth]{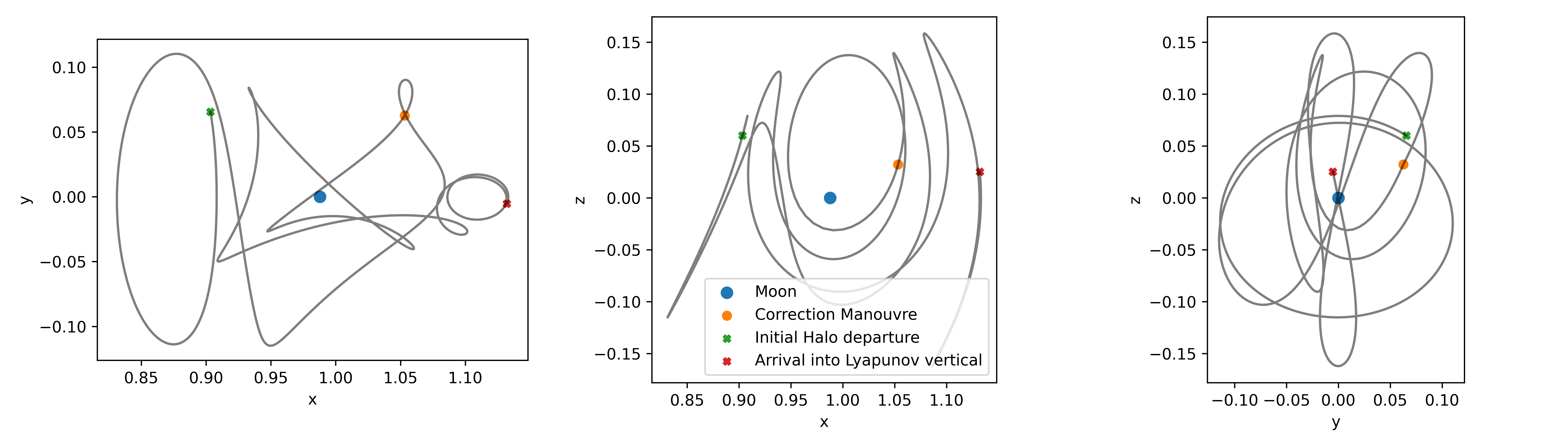}
    \caption{The reference heteroclinic connection between a Halo and a vertical Lyapunov orbit. A single impulsive correction manoeuvre is present.}
    \label{fig:heteroclinic}
\end{figure*}

\subsection{A Halo-Vertical Lyapunov heteroclinic transfer}
As a third example, we consider the circular restricted three-body problem in the Earth-Moon system. 
We start with an almost heteroclinic connection between a Halo orbit around $L_1$ and a vertical Lyapunov orbit around $L_2$: this trajectory connects the two periodic orbits but requires a small $\Delta V$ to be applied along the trajectory. 
The transfer requires a $29.31$ m/s correction maneuver after 25.83 days and reaches the vertical Lyapunov orbit along the corresponding stable manifold after 58.63 days in total. 
This trajectory was taken among those presented in the work of Tagliaferri et al. \cite{tagliaferri}, where a systematic search procedure is outlined for these cases.
However, the small mid-course position discontinuity (characteristic of the approach in \cite{tagliaferri}), was canceled here with a few correction steps. 
The state transition matrix is computed via the variational equations. Due to the chaotic nature of the system and the particularly sensitive trajectory, the STM computations have a rather large numerical instability associated that needs to be taken care of by resetting the corresponding initial conditions frequently and performing the numerical integration forward and backward up to a midpoint.
This, in connection to the high precision Taylor scheme employed \cite{biscani2021revisiting} allows to contain this intrinsic instability to a tolerable amount. 
Formally this specific case was not explicitly covered above, as it makes use of one impulse in the middle of the trajectory arc, however, the same formalism can be easily adapted to this situation. 
The results are shown in Figure \ref{fig:heteroclinic_s}, in which areas where the primer condition allows for the convenient addition of two impulses are highlighted. Several occasions exist to improve this trajectory. Notably the highest primer condition is attained at a point $t_1, t_2$ just after the mid-course correction manoeuvre. Adding there two impulses, and reoptimizing the trajectory, we get a new total $\Delta v=20.31$ m/s thus lowering significantly the previous transfer as expected.

\section{Conclusion}
In the context of multiple impulse trajectories, the classical primer vector, rooted in Pontryagin's optimal control theory, traditionally facilitates the addition or removal of impulses. In our work, we demonstrate an alternative approach, leveraging only first-order information surrounding a reference trajectory, as provided by the state-transition matrix, and without introducing adjoint variables (co-states). Our derivation aims for simplicity and clarity while achieving the same conditions. Subsequently, we utilize this new formalism to derive novel optimality conditions, covering previously intractable cases. Specifically, we construct a surrogate primer vector that allows us to assess the benefits of adding two new impulses to a single $\Delta \mathbf V$ trajectory. Although the construction involves a simple numerical optimization, it proves highly efficient and unveils the intricate structure of multiple impulse trajectories. We then successfully apply the new surrogate primer vector to three cases: a simple transfer in Keplerian dynamics, a more complex scenario involving the MSR-ERO mission in an accurate ephemeris model, and an almost heteroclinic transfer between a Halo orbit and a vertical Lyapunov orbit in the Earth-Moon system

Due to its broad derivation applicable to any multi-impulsive trajectory, our findings seamlessly extend to diverse applications. Notably, our developments offer insights into optimizing station-keeping maneuvers around Lagrange points and refining quasi-heteroclinic connections between Halo orbits. This generality underscores the versatility and potential impact of our approach across various spaceflight mechanics scenarios.

\section{Code availability}
All of the figures and quantitative results in this paper can be reproduced running the python based code distributed freely under a permissive GPLv3 license here: \url{https://gitlab.com/EuropeanSpaceAgency/mit}. Refer to this resource to have all details on the dynamical system used and the corresponding values of their paramters and initial conditions.

\bibliographystyle{ISSFD_v01}
\bibliography{references}

\end{document}